%% file: main.tex
\documentclass[peerreview]{IEEEtran}

\author{\IEEEauthorblockN{Steven W. D. Chien$^1$, Stefano Markidis$^1$, Vyacheslav Olshevsky$^1$, \\Yaroslav Bulatov$^2$, Erwin Laure$^1$, Jeffrey S. Vetter$^3$\\} 
	\IEEEauthorblockA{
		$^1$ KTH Royal Institute of Technology, Stockholm, Sweden \\
		$^2$ South Park Commons, San Francisco, USA\\
		$^3$ Oak Ridge National Laboratory, Oak Ridge, USA 
	}
}

\usepackage{amsmath,amssymb,amsfonts}
\usepackage{algorithmic}
\usepackage{graphicx}
\usepackage{textcomp}
\usepackage{xcolor}
\usepackage{listings}
\definecolor{strings_color}{rgb}{1.0,0.0,0}
\definecolor{comments_color}{rgb}{0.2,0.2,0.2}
\usepackage{hyperref}

\begin{document}

\title{TensorFlow Doing HPC \\
{\large An Evaluation of TensorFlow Performance in HPC Applications}}

\maketitle

\begin{abstract}
	TensorFlow is a popular emerging open-source programming framework supporting the execution of distributed applications on heterogeneous hardware. While TensorFlow has been initially designed for developing Machine Learning (ML) applications, in fact TensorFlow aims at supporting the development of a much broader range of application kinds that are outside the ML domain and can possibly include HPC applications. However, very few experiments have been conducted to evaluate TensorFlow performance when running  HPC workloads on supercomputers. This work addresses this lack by designing four traditional HPC benchmark applications: STREAM, matrix-matrix multiply, Conjugate Gradient (CG) solver and Fast Fourier Transform (FFT). We analyze their performance on two supercomputers with accelerators and evaluate the potential of TensorFlow for developing HPC applications. Our tests show that TensorFlow can fully take advantage of high performance networks and accelerators on supercomputers. Running our TensorFlow STREAM benchmark, we obtain over 50\% of theoretical communication bandwidth on our testing platform. We find an approximately $2\times$, $1.7\times$ and $1.8\times$ performance improvement when increasing the number of GPUs from two to four in the matrix-matrix multiply, CG and FFT applications respectively. All our performance results demonstrate that TensorFlow has high potential of emerging also as HPC programming framework for heterogeneous supercomputers.
\end{abstract}


\input{introduction}
\input{background}
\input{methods}
\input{evaluation}
\input{results}
\input{related-work}
\input{conclusion}

\section*{Acknowledgment}
Funding for the work is received from the European Commission H2020 program, Grant Agreement No. 801039 (EPiGRAM-HS). Experiments were performed on resources provided by the Swedish National Infrastructure for Computing (SNIC) at PDC Center for High Performance Computing and HPC2N.

\bibliographystyle{IEEEtran}
\bibliography{main}

\end{document}

%% file: introduction.tex
\section{Introduction}
\label{sec:introduction}
TensorFlow is a fast growing open-source programming framework for numerical computation on distributed systems with accelerators. It was originally developed by Google and made open-source in November 2015. Since then, TensorFlow grew at a considerable pace counting on the work of thousands of active developers spreading across the world. In 2017, the TensorFlow code repository recorded more than a thousand commits per month, making it one of the fastest growing open-source software projects.

TensorFlow lowers the difficulty of programming accelerators in distributed environments by avoiding HPC low-level programming interfaces, such as CUDA, OpenCL and MPI. These concepts are essential for programming any HPC application. As such, TensorFlow enables the use of cloud systems with accelerators, such as Volta GPUs~\cite{markidis2018nvidia}, TPUs~\cite{jouppi2017datacenter} and current pre-exascale supercomputers, such as Summit and Sierra supercomputers, easily accessible and usable.

While TensorFlow was originally designed after Google's DistBelief~\cite{dean2012large} for solving Machine Learning (ML) problems, the ultimate goal of TensorFlow is to solve a much broader range of numerical problems. In fact, the TensorFlow code repository webpage states ``\emph{TensorFlow is an open source software library for numerical computation}''. However, few examples of how to use TensorFlow for solving generic numerical problem on HPC systems exist, making it difficult to evaluate the impact of TensorFlow framework outside of the ML domain. In addition, TensorFlow is mostly deployed on local workstations or on the cloud, thus few performance data of TensorFlow running on supercomputers exist. This work aims to fill this gap by designing and developing a set of HPC applications and measuring their performance on supercomputers with accelerators. 

Motivated to understand the overall impact of TensorFlow in HPC field, we design and implement four main benchmarks and computational kernels which are widely used by the HPC community: STREAM, matrix-matrix multiply, a Conjugate Gradient~(CG) linear solver and a Fast Fourier Transform~(FFT). In the formulation of such algorithms, we follow a ``data-driven'' approach similar to that of a ML pipeline, where we tile the input matrices or arrays in smaller tiles and offload calculation to accelerator using the distributed TensorFlow. We develop a TensorFlow adaptor for the Slurm resource manager~\cite{yoo2003slurm} to easily deploy TensorFlow applications on supercomputers. In addition, we run scaling and performance tests of our HPC applications on two supercomputers with GPUs. Our main contributions are to show that \emph{i)}~By using TensorFlow, application developers can easily develop parallel applications capable of using supercomputers with GPUs without having to deal with MPI, CUDA or OpenCL. \emph{ii)}~TensorFlow applications, designed to follow a ``data-driven'' approach, show relatively good performance and scaling as the number GPUs increases. Overall, we show that potential of TensorFlow uptake by the HPC community is high, especially when considering the current difficulty of programming large supercomputers with accelerators.

The paper is organized as follows. We first provide an overview of the TensorFlow programming environment, focusing on HPC aspects in Section~\ref{sec:environment}. We follow by describing how to deploy a TensorFlow application on supercomputers in Section~\ref{sec:slurm} and present the HPC applications using TensorFlow in Section~\ref{sec:hpc-application}. The experimental set-up is described in Section~\ref{sec:experiment} and we present the performance results in Section~\ref{sec:results}. We describe previous and related work on TensorFlow programming environment and other ML frameworks in Section~\ref{sec:related-work} and finally summarize and discuss the results, outlining future work in Section~\ref{sec:conclusion}.

%% file: background.tex
\section{An Overview of TensorFlow}
\label{sec:environment}
TensorFlow's name is inspired by the idea of tensors propagating through computation graph. As stated in the GitHub TensorFlow repository, in a data flow graph, ``\emph{the graph nodes represent mathematical operations, while the graph edges represent the multidimensional data arrays ({\bf tensors}) that {\bf flow} between them}''. All data in TensorFlow are represented as n-rank tensors that are in practice n-dimensional arrays of basic data-types: a 0 dimensional tensor represents a scalar, a one-dimensional tensor represents a vector, a two-dimensional tensor represents a matrix and higher dimension tensors corresponds to higher dimension matrices. Tensor in TensorFlow has two properties: data type and shape which correspond to the number of dimensions and size for each dimension. All kinds of Tensors in TensorFlow, such \textsf{\small{tf.placeholder}}, \textsf{\small{tf.constant}} and \textsf{\small{tf.SparseTensor}}, are immutable with the exception of \textsf{\small{tf.Variable}}.

TensorFlow uses dataflow computing paradigm: computation of tensors is expressed in terms of the dependencies between individual operations~\cite{tensorflow-computation-graph}. In TensorFlow, a dataflow graph is first defined, then the dataflow graph or parts of the graph are executed across a number of local and remote devices through a TensorFlow session. The major advantages of using dataflow computing in HPC context are:

\begin{itemize}
\item With dependencies between operations known before running the DAG, TensorFlow can identify operations that can execute in parallel.
\item  TensorFlow uses information of dataflow between operations to perform parallelization.
\item  TensorFlow can use information of the dataflow graph to optimize execution, for instance merging subsequent operations to avoid data movement.
\end{itemize}

In addition, the use of dataflow graph allows portability because dataflow graph is a language-independent representation of the program. For instance, the graph can be constructed through one language binding, e.g. in Python, and be reopened later in a C++ code.

The current version of TensorFlow uses a deferred execution model (called \emph{Graph mode} in TensorFlow) where the graph is constructed and operations are subsequently executed as soon as a TensorFlow Session is instantiated and executed. We note that TensorFlow also supports \emph{eager execution} that follows an imperative style and it will likely become the default execution mode in future releases of TensorFlow.

\begin{figure}[t]
	\begin{center}
		\includegraphics[width=0.4\linewidth]{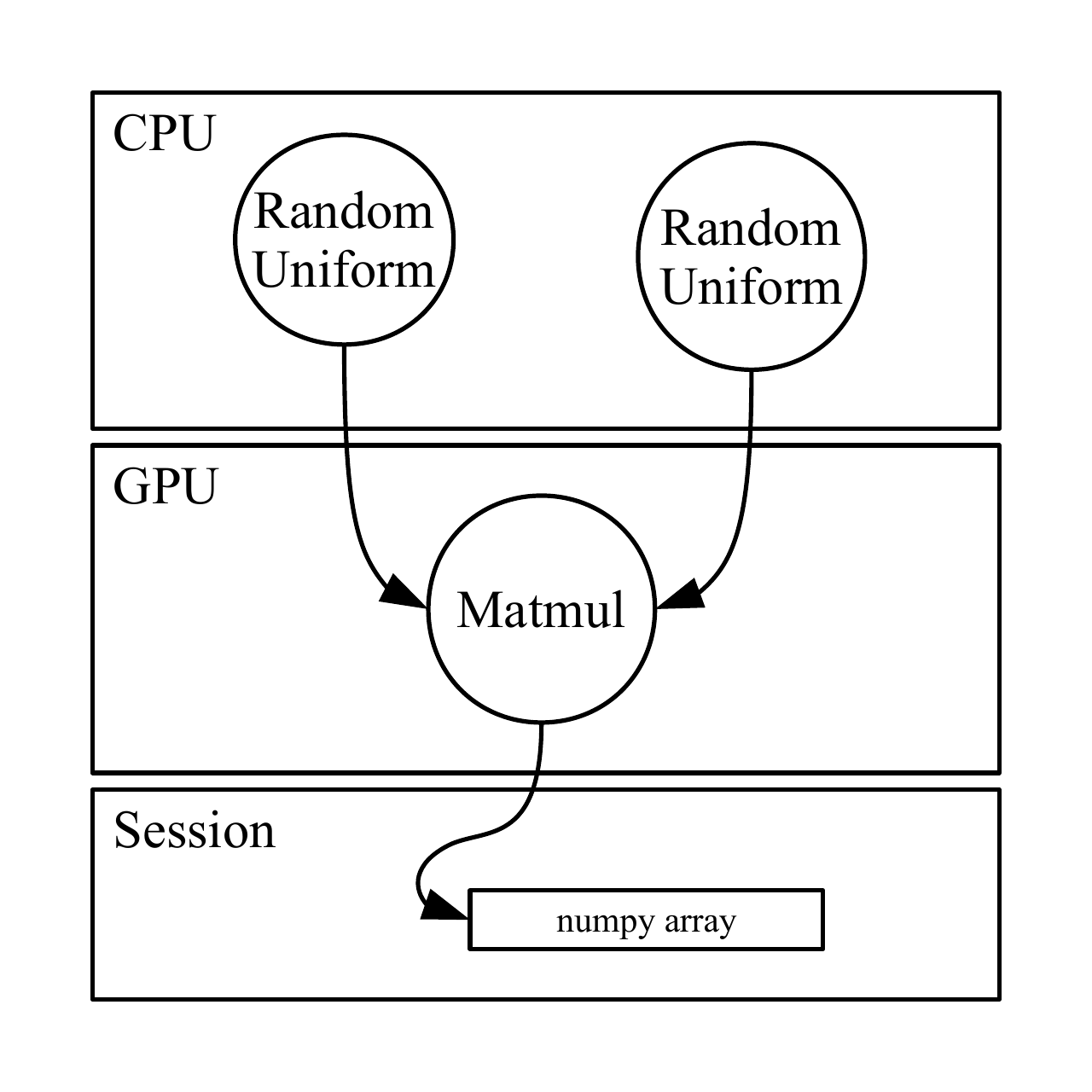}
		\caption{Computation graph and execution through session for code snippet generated by Listing~\ref{lst:device-placement}. When invoked, two matrices are randomly generated on CPU then transferred to the GPU for multiplication. The final result is transferred to the client through the session as a \emph{Numpy} array.}
		\label{fig:device-placement}
	\end{center}
\end{figure}

Operations on a computation graph can be deployed through manual or automatic pinning with \textsf{\small{tf.device()}}. At the moment, the two available device types are \emph{cpu} and \emph{gpu}. Support for other devices can be implemented through designing a respective computation kernel and a wrapper API. If no device is specified, a \emph{simple device placement} is used: if an operation supports both CPU and GPU execution, GPU devices will be chosen. In case when multiple GPUs are available, the first GPU will be chosen. Another approach is \emph{soft device placement}. When an operation is pinned to a device with no supporting computation kernel, it can be automatically pinned to another device with a supporting kernel instead. In a heterogeneous environment with two GPUs, one CPU and no device specification, a matrix multiplication \textsf{\small{tf.matmul()}} will be pinned to the first GPU (GPU with device number zero).

\begin{lstlisting}[caption=Computing a matrix multiplication where A and B are generated in CPU while multiplication is done on GPU, label={lst:device-placement}]

import tensorflow as tf

g = tf.get_default_graph()
with g.as_default():
  with tf.device('/cpu:0'):
    a = tf.random_uniform(shape=[3, 3],
                          dtype=tf.float32)
    b = tf.random_uniform(shape=[3, 3],
                          dtype=tf.float32)
  with tf.device('/gpu:0'):
    c = tf.matmul(a, b)

with tf.Session(graph=g) as sess:
  ret_c = sess.run(c)
\end{lstlisting}

Listing~\ref{lst:device-placement} illustrates an example of a matrix multiplication where two random matrices are generated on a CPU while the multiplication is performed on GPU. The code snippet generates a graph similar to that of in Fig.~\ref{fig:device-placement}. Data movement between host and GPU memory is automatically handled and scheduled by the runtime without the user needing to handle any allocation.

\subsection{Distributed Model}

Before developing a TensorFlow distributed application, it is important to master few TensorFlow concepts, such as TensorFlow server, cluster, tasks and jobs, and to have it clear that TensorFlow uses a Remote Procedure Call (RPC) service-client model.

To run TensorFlow in a distributed environment, we need first to set-up a TensorFlow cluster. A TensorFlow cluster consists of multiple TensorFlow servers, also called tasks. A TensorFlow job is a named set of tasks that have common function. Typically, two kinds of jobs are defined: the \emph{parameter server} or \emph{ps} jobs host nodes that store and update variables; worker jobs are responsible for compute-intensive tasks and host stateless variables.
\begin{lstlisting}[caption=A TensorFlow Cluster specification that comprise of two jobs: parameter server and workers. Each job has a list of addresses and port numbers to TensorFlow servers., label={lst:cluster-spec}]

cluster = tf.train.ClusterSpec(
{ 'ps':     ['t01n01:8888'],
'worker': ['t01n02:8888', 't01n03:8888'] })
\end{lstlisting}
The Cluster is set-up by using a cluster specification, defining jobs and tasks. Listing~\ref{lst:cluster-spec} shows how cluster specification comprising two jobs with one and two tasks respectively looks like. A dictionary of jobs is specified in a cluster specification and each job consists of a list of addresses and port numbers to TensorFlow servers that are responsible for that particular task.

A \textsf{\small{tf.train.Server}} object then needs to be created with the cluster specification as an argument so that communications to other servers can be established. TensorFlow allows more than one task to be launched per node. If the servers use different GPUs in a node, we need to ensure that the respective GPUs are exposed. This can be done through configuring CUDA environment variable. If more than one server are using one GPU, we need to ensure that the two tasks share the GPU memory. In fact, by default each TensorFlow instance takes all the GPU memory. For this reason, configurations on how memory is allocated will need to be explicitly specified in this case.

The RPC service-client model is a popular model to design data-driven applications~\cite{kleppmann2017designing}. RPC is simple and usually platform and language independent. RPC protocols are typically designed on top of message passing systems. In TensorFlow, the Google RPC library\footnote{https://grpc.io} is used. gRPC is a library that provides language independent RPC service between clients and server. It uses Protocol Buffer\footnote{https://developers.google.com/protocol-buffers/docs/overview} as underlying message serialization and exchange. ProtoBuf is a language and platform independent serializable data structure developed by Google. Apart from communication, it is widely used by TensorFlow such as for representation and serialization of dataflow graph.

TensorFlow supports two parallelism models:
\begin{itemize}
 \item { \bf Model Parallelism}: The computational graph is split across different devices such as in Fig.~\ref{fig:device-placement}.
 \item { \bf Data Parallelism}: A graph is replicated for each task and they individually execute their own graph. According to computation results they update parameters on the parameter server. In each computational step, each task fetches different data for computation.
 \end{itemize}
In this work, we focus on data parallelism, where each task individually executes a graph with different data source. The two main ways of handling data movement is through the Dataset API, where data are loaded, pre-processed and prefetched as tensors through an input pipeline such that data is ready for immediate consumption~\cite{chien2018characterizing}; another method is based on the Queue API, where tasks can push tensors into or extract tensors from a shared queue.

\subsection{The TensorFlow Programming Environment}
\begin{figure}[t]
	\begin{center}
		\includegraphics[width=0.7\columnwidth]{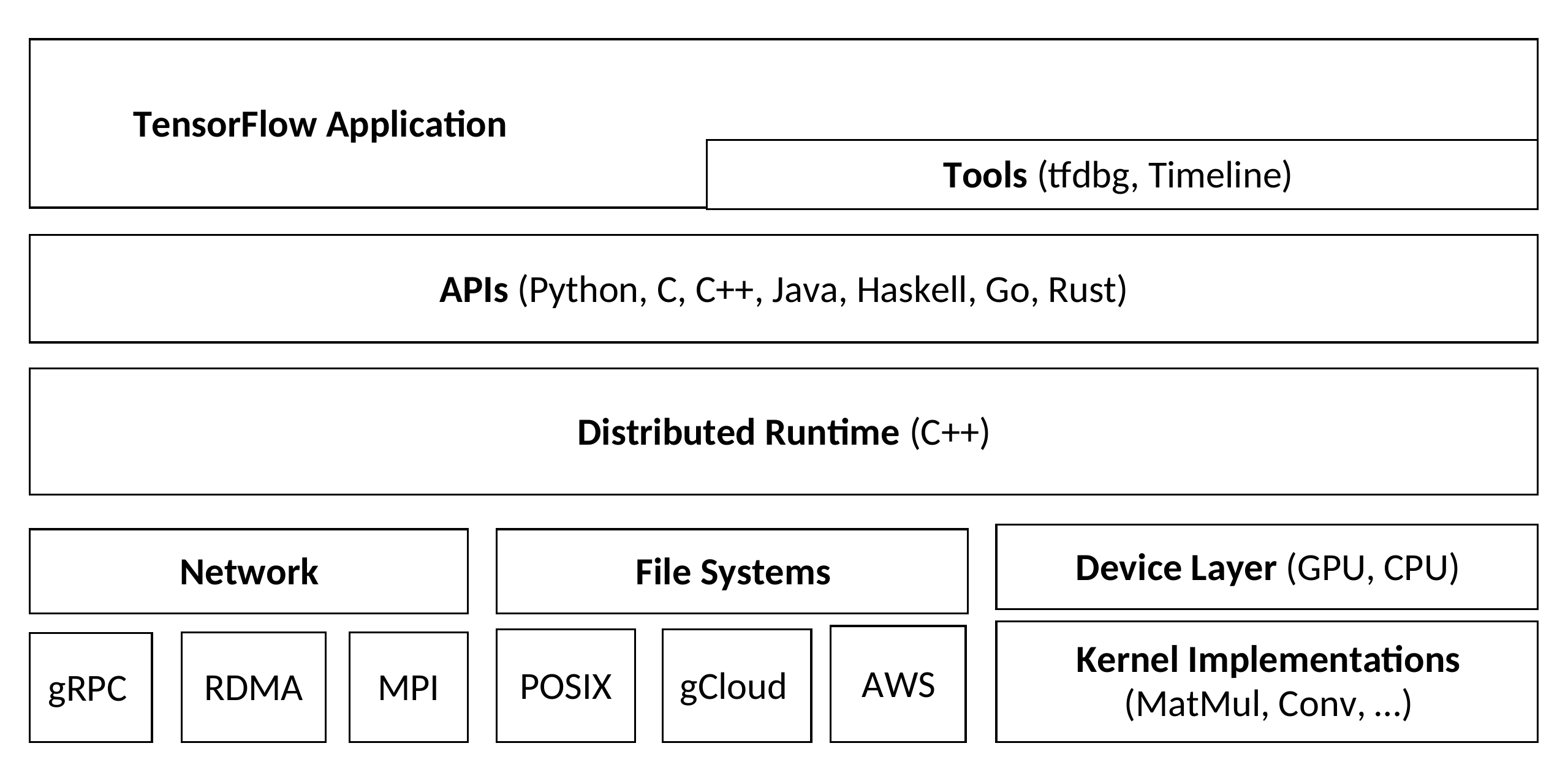}
		\caption{TensorFlow software stack, adapted from Ref.~\cite{tensorflow}.}
		\label{fig:software-stack}
	\end{center}
\end{figure}
One of the main strengths of TensorFlow software framework is that it combines high-level APIs with a lower level distributed C++ runtime system. TensorFlow's software stack is illustrated in Fig.~\ref{fig:software-stack}. It provides a set of high level APIs, below which, is supported by a runtime written in C++. The runtime uses different low level modules, such as device execution kernels, file I/O and network communication modules to support execution. To build a graph, a client program executes different operation construction APIs provided by the TensorFlow API. It finally establishes a session to their own server to run the graph. While TensorFlow provides a variety of APIs, among which C, Java, Haskell, Go and Rust APIs, the most used API is the Python API since it has complete support for all the functionalities and can be combined with other Python modules for numerical computation, such as \emph{Numpy} and \emph{scikit-learn}. As most of the TensorFlow users, we also use the Python API in this work.

The C++ runtime is the core of TensorFlow and provides HPC implementations for handling communication across the network, operations on accelerators, and I/O to the devices. Particurlarly important for HPC, apart from gRPC, TensorFlow supports the use of InfiniBand Verbs RDMA which was contributed by Yahoo; and MPI for data transfer. When these protocols are used, gRPC is typically still responsible for administrative purposes such as to establish initial connections.

One feature that might be of interest to HPC application users is its ability to take checkpoints and restart from an execution state. When a snapshot is taken, the graph structure, checkpoint identification and actual data stored in variables are recorded and stored as files~\cite{chien2018characterizing}. Computation can be restarted later by restoring the graph structure and data stored in graph variables. Snapshot taking and restoration can be controlled through the TensorFlow API.

\begin{figure}[t]
	\begin{center}
		\includegraphics[width=0.7\columnwidth]{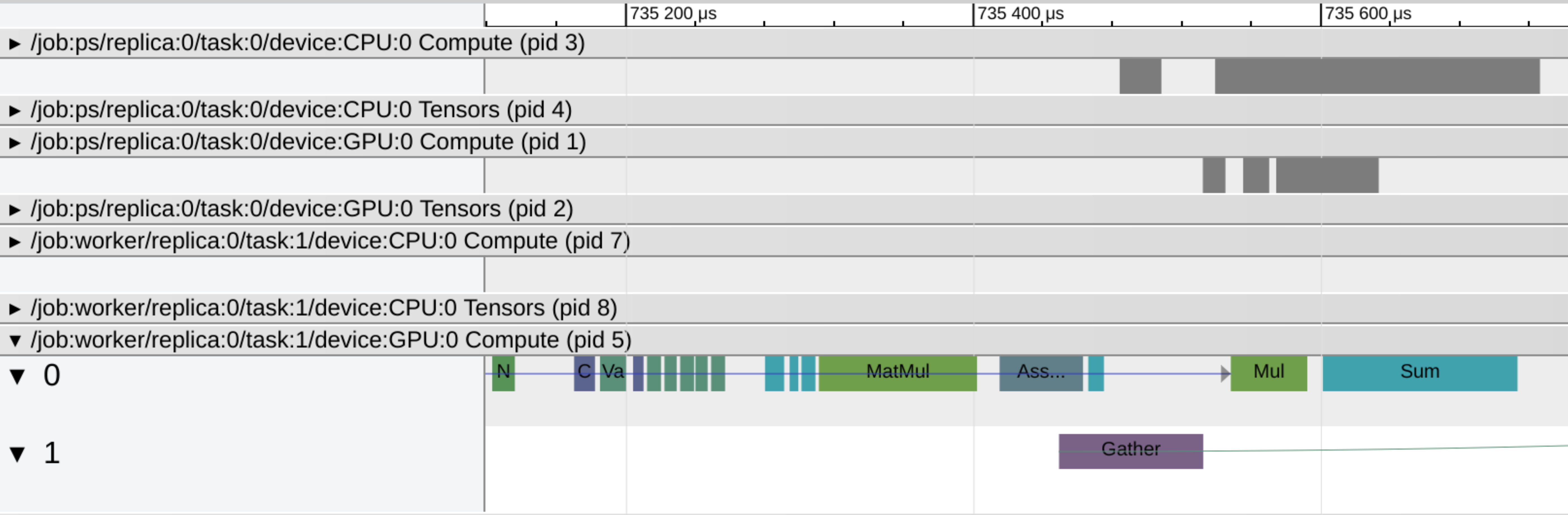}
		\caption{Execution TensorFlow Timeline of a particular stage of our CG solver. The individual time lines of a device show parallel execution of models.}
		\label{fig:timeline}
	\end{center}
\end{figure}
TensorFlow provides developers tools to perform runtime performance analysis. One example is TensorFlow Timeline. During session execution, session run metadata which consists of information on such as CUDA stream, and graph execution statistics are collected and can be visualized in form of a time line. Fig.~\ref{fig:timeline} shows an example of a time line analysis of an execution of our CG solver application. \emph{tfdbg} is an interactive debugger which gives a command line interface and is mainly for runtime debugging inside a computation graph. With \emph{tfdbg} it is possible to inspect contents of tensors and computation during execution.

%% file: methods.tex
\section{Running TensorFlow on Supercomputers}
\label{sec:slurm}
One current limitation of running distributed TensorFlow on HPC systems is the lack of integration of TensorFlow and job schedulers, such as Slurm on the systems. TensorFlow introduces ``Cluster Resolver" that is responsible for fetching job and computing node configurations thus allow easy launching of tasks. Resolvers for launching job on TPUs and Google Cloud Service are already provided. In this work, we have developed a TensorFlow Resolver to automatically specify TensorFlow Cluster specifications and for launching tasks on a supercomputer. 

Jobs in HPC systems are commonly managed by workload managers such as Slurm through a batch queuing system and the computing nodes allocated are not known in advance. In addition, these computing nodes are not directly accessible to end-users. Launching and placement of tasks are handled by the workload manager. For this reason, we extended the \textsf{\small{tf.contrib.cluster\_resolver}} module to support cluster specification resolution for jobs launched by Slurm. \textsf{\small{tf.contrib.cluster\_resolver}} is an object that automatically generates a cluster specification with given parameters. Our implementation supports homogeneous job allocation with default Slurm plane distribution. The object is created by specifying a list of jobs, number of tasks on each node and how GPUs on nodes are allocated to tasks. The Resolver reads a list of hosts through Slurm's \textsf{\small{scontrol}} command and determine how jobs and tasks will be distributed. GPUs will also be exposed automatically to processes in case multiple TensorFlow instances are running on the same node.

\section{From ML to HPC Applications}
\label{sec:hpc-application}
TensorFlow has been designed with ML data-driven workloads in mind. In typical ML frameworks, data, such as images, are ingested at a fast pace. Data is preprocessed via an input pipeline on host device and moved to accelerator memory where computation is performed. Finally, parameters are updated according to results of computation. With training data-driven ML workloads, the amount of data being moved to the GPU is typically large, thus performance is often limited by communication bandwidth. 

To achieve high-performance with TensorFlow on supercomputers, HPC application algorithms need to be reformulated as data-driven problem. We choose to design and implement four main applications which are widely used in the HPC Community. When developing our applications in TensorFlow, we also use additional modules, such as \emph{Numpy}, for merging and other auxiliary operations.

{\bf A TensorFlow STREAM.} STREAM is among the most famous HPC benchmarks. It measures the sustained bandwidth to memory systems. We implement a STREAM~\cite{stream} like application to study the communication performance between two GPUs, also CPUs located on two computing nodes. We create a simple TensorFlow cluster with two tasks, a parameter server and a worker on the two nodes. A vector of floating point numbers with a specified size is created on each GPU belonging to the two computing nodes. We create an \textsf{\small{assign\_add}} operation which pushes the vector located on the GPU of the worker to parameter server and performs addition with the vector located on the GPU of parameter server. The application supports the use of gRPC, MPI and InfiniBand Verbs as underlying transfer protocol. To measure the estimated cost of transfer, we open a session to the worker and invoke the assign add operation through the session. Time used for invoking the operation can be conceived as the estimated cost of transfer between the two devices. When invoking an operation through a session, TensorFlow automatically returns the evaluated value from the graph in form of a \emph{Numpy} array. In the case of bandwidth measure, this means additional data transfer. For this reason, we explicitly specify not to have the evaluated value returned to the Python session.

\begin{figure}[t]
	\begin{center}
		\includegraphics[width=0.7\linewidth]{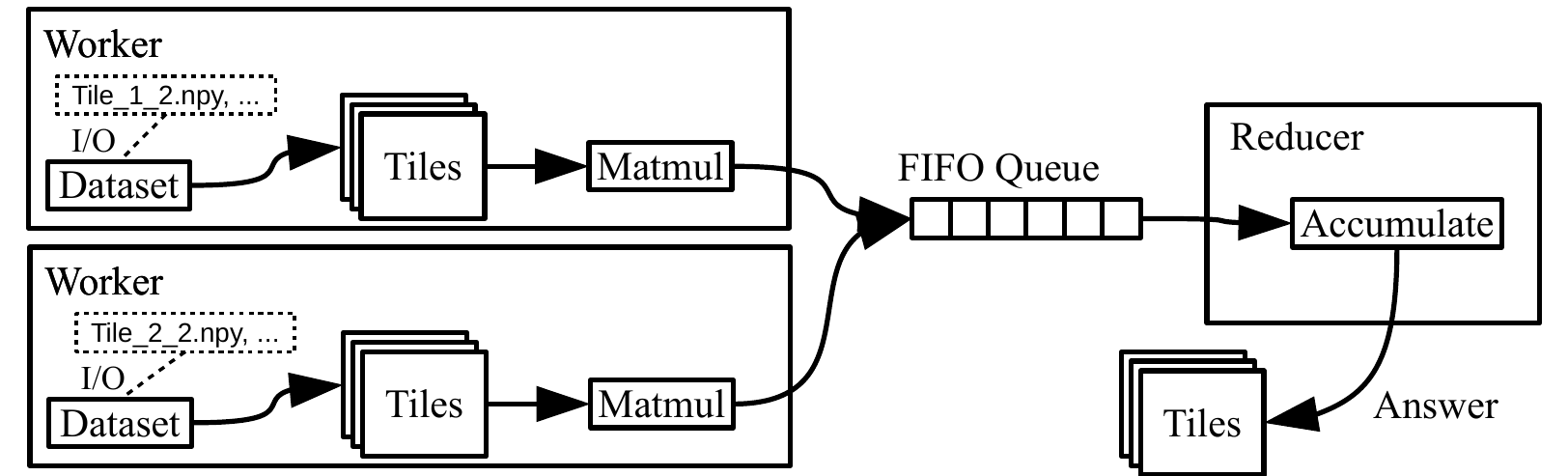}
		\caption{A simplified illustration of tiled matrix multiplication application. Blocks of two matrices to be multiplied are fetched into a dataset with an index of which tile the multiplication result should be accumulated to. Workers continuously obtain tiles from dataset, perform matrix multiplication of the tiles and push results to one or more queues with the target index. One or more reducers executed inside the session on the reduce job: collecting multiplication results from its respective queue and accumulate to the respective \emph{Numpy} array.}
		\label{fig:matmul-design}
	\end{center}
\end{figure}

{\bf Tiled Matrix Multiplication.} We implement matrix-matrix multiplication of large matrices similar to that in Fig.~\ref{fig:matmul-design}. With ``large", we mean that the whole matrices cannot be stored in the memory of one GPU. We reformulate computation similar to that of a ML training pipeline. To parallelize computation, we use a standard technique used in GPU programming, called tiled matrix-matrix multiply: two large matrices are divided into equally sized tiles. These tiles are stored and loaded as \emph{Numpy} arrays. We create a dataset which gives a list of indexes of tiles to be multiplied and their resulting index. The list is shared by workers and they individually load these tiles from files. These tiles are loaded to GPU memory and multiplication is performed. The resulting tile is pushed into a First-In-First-Out queue of a reducer. The reducers extract these tiles and locally accumulate them to their respective targets. The algorithm is embarrassingly parallel and resembles a map-reduce paradigm, where tiles are multiplied and sent to the respective queues (\emph{map}) and finally accumulated and stored by reducers (\emph{reduce}). Furthermore, the input pipeline resembles a typical ML training data input pipeline, where samples (tiles) are loaded and sent through a computation pipeline (matrix multiplication) on GPU. In our implementation all computations are performed in single precision (32-bit).

{\bf Conjugate Gradient Linear Solver.} Linear solvers are widely used to solve Partial Differential Equations (PDE) that arise in engineering, physics and chemistry problem. CG is an iterative linear solver for semi-definite positive matrices. The parallelizable elements in the algorithm include matrix vector multiplication and dot product. Matrix vector multiplication can be independently computed by horizontally splitting the matrix and distributing to worker such that each workers are responsible for one part of the multiplication. For example, a $8192 \times 8192$ matrix can be split into $2048 \times 8192$ blocks for four workers such that they can compute a share of the result vector. The final result can be obtained by merging the vectors. A dot product of two vectors can be computed in parallel by distributing subset of vectors to workers. Workers compute their share of dot product and the final result can be obtained by summing all the partial dot products. Our approach of expressing the algorithm resembles similarities to what was presented in Ref.~\cite{10.1007/978-3-642-36424-2_15}, where the algorithm is presented as a dependency specified dataflow graph for FPGAs.

\begin{figure}[t]
	\begin{center}
		\includegraphics[width=0.5\linewidth]{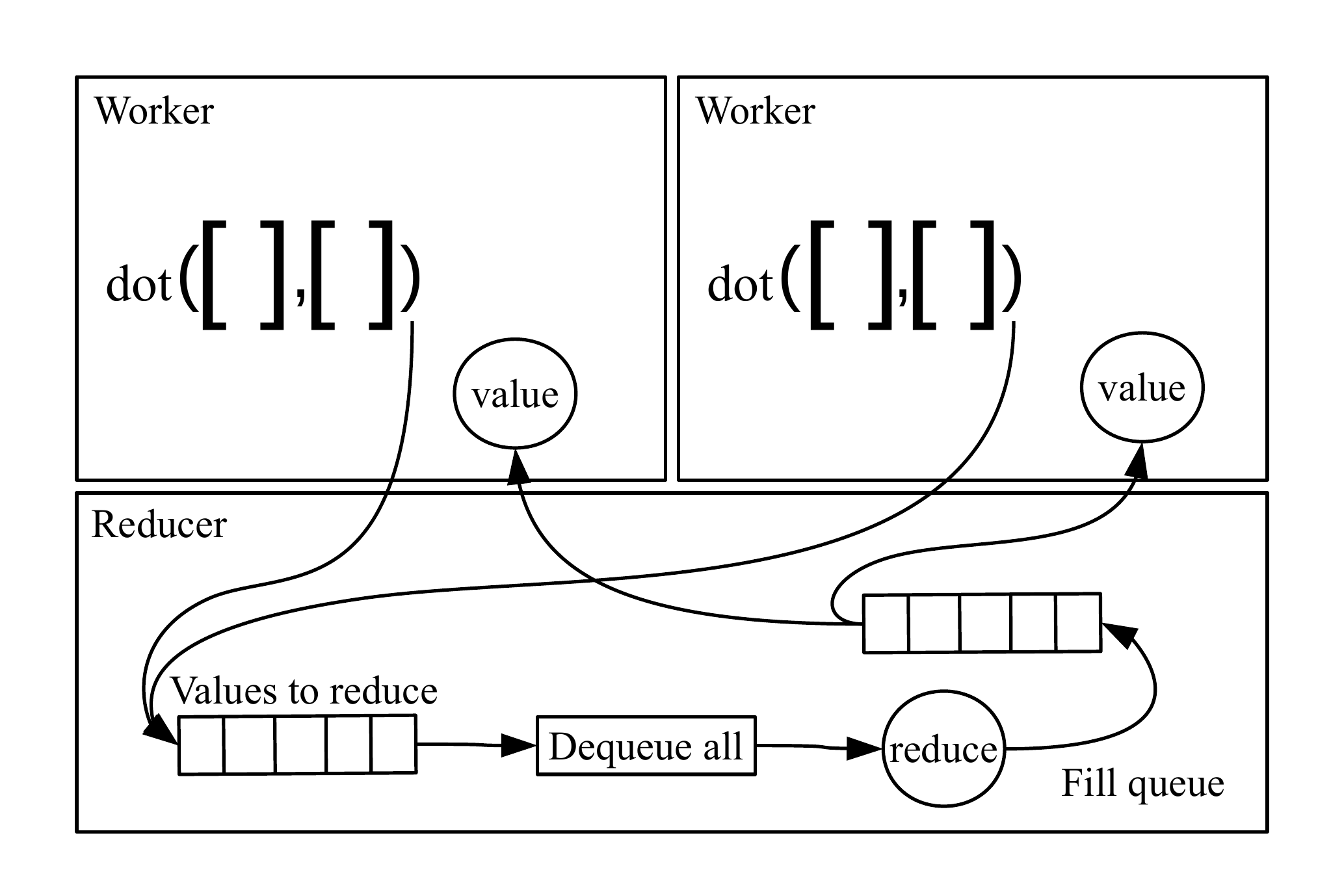}
		\caption{A simplified illustration of data-driven reduction workflow between workers in the CG application. Workers participate in a reduction by sending their respective values into the incoming queue of a reducer, then wait to dequeue from the outgoing queue. The reducer extracts values from the incoming queue and preforms requested operation. The reduced value is distributed to workers through the outgoing queue. A number of copies equivalent to the total number of workers will be pushed into the queue. Each worker which is waiting to dequeue obtains a copy of the reduced value and proceeds with own execution.}
		\label{fig:cg-reduce-design}
	\end{center}
\end{figure}

One limitation of TensorFlow is that computation graphs, which are represented as ProtoBuf, cannot exceed two gigabyte in size. This imposes significant limitation to problem size. This can be overcome by creating uninitialized variables and explicitly update them inside a session through dictionary feeding. For example, one can represent the results of calculation of a loop as a single unrolled TensorFlow graph. However, this can exceed the two gigabyte limitation on the size of computation graph. Instead, one can represent only the body of the loop as the computation graph, and use TensorFlow persistent storage (\emph{tf.Variable}) to keep the state between iterations. For this reason, we introduce data locality by loading and storing tiles from a matrix directly to workers to be reused every iteration.

We reformulate the synchronization and reduction steps with data-driven approach. When performing distributed synchronous learning, TensorFlow provides a \textsf{\small{SyncReplicasOptimizer}} where updates to a variable will be collected from workers and averaged before being applied. It uses a token queue as an implicit barrier and distributes updated step counters to worker after variable updates by populating a queue, where workers dequeue from. We follow this approach and implement a reducer where workers push updates to the reducer through a queue and extract the reduced value from the reducer through another queue. As illustrated in Fig.~\ref{fig:cg-reduce-design}, two queues are created for each reduction step in the algorithm: one for workers sending in partial values for reduction operation and the second for distributing reduced value back to workers.

For example, each worker individually computes intermediate values and pushes them into the incoming queue of reducer and enters a blocking wait for output queue. The reducer collects all scalars being sent by the workers, performs computation and populates an outgoing queue with the updated scalar. Workers extract the updated scalar and proceed to next stage of computation. In that sense, data are driven continuously between workers and reducers. Our solver performs computation in double precision (64-bit).

\begin{figure}[t]
	\begin{center}
		\includegraphics[width=0.5\linewidth]{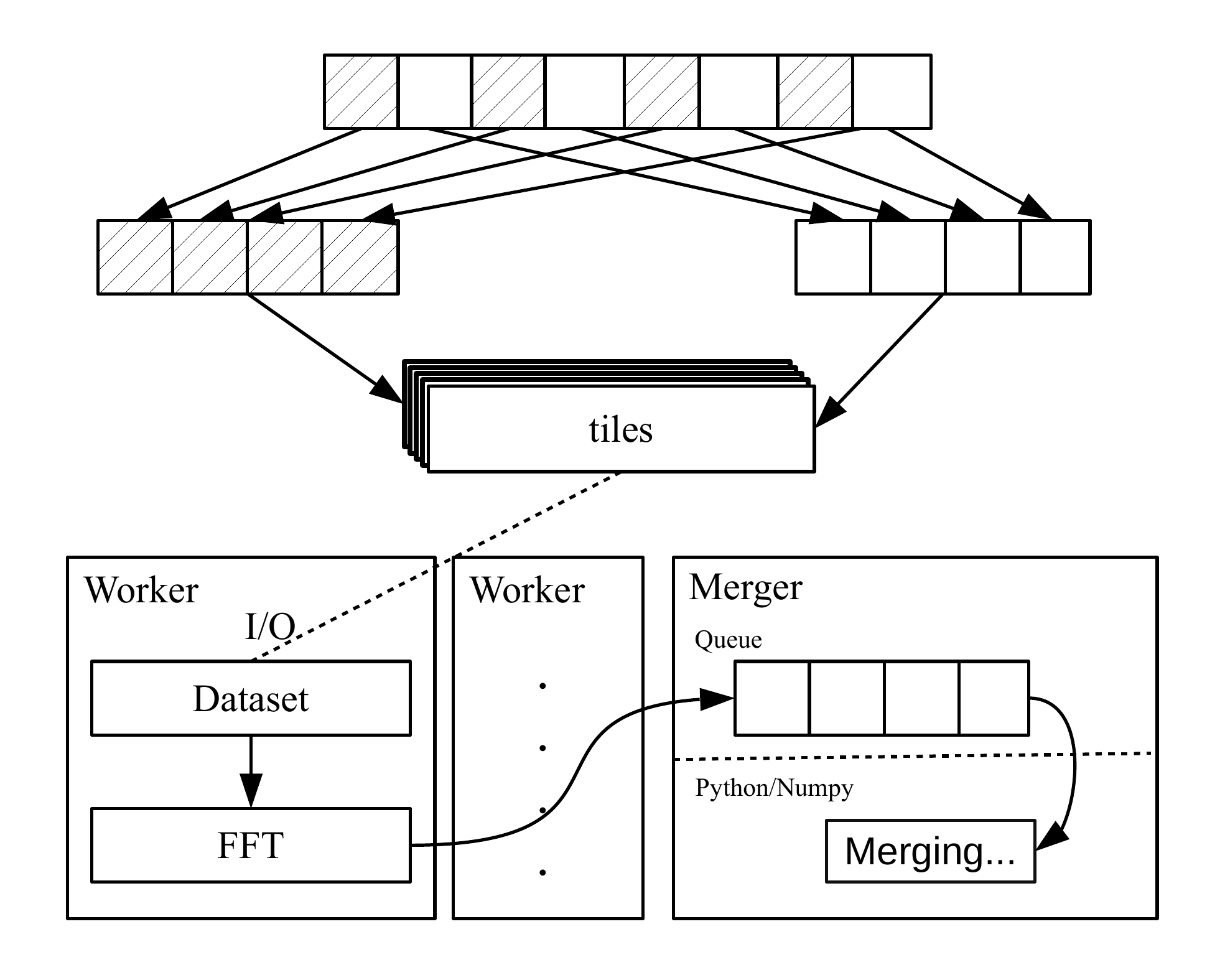}
		\caption{A simplified illustration of data-driven 1D FFT. The input vector is split into interleaving tiles and stored. Each worker load their responsible tiles and perform FFT. The result from FFT is pushed into a result queue and a merger continuously collect the resulting tiles and their index. Once all results are collected the merger computes twiddle-factors and performs merging locally with Python.}
		\label{fig:fft-design}
	\end{center}
\end{figure}

{\bf Fast Fourier Transform (FFT).} FFT is the algorithm for performing Fourier transform on a discrete signal of size N in O(N log(N)) operations. FFT is a key algorithm for computational physics that is widely used in signal processing, spectral analysis and for solving PDEs. We implement the Cooley-Tukey 1D FFT algorithm as illustrated in Fig.~\ref{fig:fft-design} where a vector is split into interleaving smaller chunks and FFT is performed over each chunk, and later merged. As in the matrix multiplication algorithm, these tiles with interleaving elements are stored in files and are loaded individually by workers which are responsible for their share of work. They perform a 1D FFT on those vectors and push the index and result to the merger's queue. The merger extracts the tiles and puts them back in place to the original vector. Once all the small vectors are collected, it computes twiddle-factors and performs merging locally with Python. We implement the solver in complex double precision (128-bit).

%% file: evaluation.tex
\section{Experimental Environment}
\label{sec:experiment}
TensorFlow has been specifically designed to run on distributed systems with GPUs. For this reason, we run our applications on two supercomputers, both equipped with GPUs:

\begin{itemize}
	\item {\bf Tegner} is a GPU cluster. Each node has two Intel E5-2690v3 Haswell processors with 512 GB of RAM. Depending on the node used each node has either one NVIDIA Quadro K420 or one K80 GPU through PCI-E. They provide 1GB and 24GB of RAM respectively. In case of K80, each card contains two GK210 GPU engines with 12 GB of RAM. The parallel file system in use is Lustre and operating system is CentOS 7.4.1708. The nodes are connected by EDR InfiniBand network. We compiled TensorFlow 1.11 with support of Python 3.6, NVIDIA CUDA, OpenMPI and InfiniBand. The versions in use are CUDA 9.1, cuDNN 7.0.5, OpenMPI 3.0 and the compiler used is GCC 6.2.0.
	
	\item {\bf Kebnekaise} is another GPU cluster. Each node has two Intel Xeon E5-2690v4 processors with 128 GB of RAM. There are two types of GPU nodes: K80 nodes or V100 nodes through PCI-E. Each K80 node has either two to four Nvidia K80 GPUs each containing two GK210 GPU engines with 12 GB of RAM or two NVIDIA V100 GPUs with 16GB of RAM. Each V100 node has two Nvidia V100 GPUs with 16 GB of RAM. The GPUs are connected through PCI-E. The operating system is Ubuntu Xenial (16.04 LTS). The nodes are connected by FDR InfiniBand network. We compiled TensorFlow 1.11 with support of Python 3.6, CUDA, OpenMPI and InfiniBand. The versions in use are CUDA 9.2.88, cuDNN 7.1.4.18 and OpenMPI 3.1.1. All tests are executed with distributed TensorFlow with InfiniBand verbs enabled. The compiler used is GCC 7.3.0.
\end{itemize}

On Tegner, we use both nodes that contain K420 and K80 GPUs. For nodes with K420, we run one instance of TensorFlow; for nodes with K80, we run two instances of TensorFlow per node and expose one GPU engine for each instance. On Kebnekaise, we use nodes that contain two K80s and V100. For K80 nodes, we run four instances of TensorFlow per node and expose one engine to each instance; similarly, for nodes with V100, we run two instances of TensorFlow per node and exposes one GPU for each instance. The number of processes running on each node is summarized in Table.~\ref{table:proc-per-node}. We also show the available memory of the GPU hosted on the node. We note that in subsequent text when we refer to K80 GPU, we refer to one GK210 GPU engine.

For each experiment, we repeat the test from five to 10 times and report the median result. We observe a minimal performance variability in all the performance tests, except for matrix-multiply on Kebnekaise K80 and FFT on Tegner with two GPUs. The applications used in this paper are hosted in a code repository\footnote{\url{https://github.com/steven-chien/tensorflow-solvers.git}}.

\begin{table}[t]
	\centering
	\caption{Number of instance of TensorFlow per node for different type of nodes in our testing platforms.}
	\begin{tabular}{|c|c|c|}
		\hline
		Type of Node    & GPU Memory      & No. processes per node \\
		\hline
		Tegner K420     & 1GB             & 1                    \\
		\hline
		Tegner K80      & 12GB $\times 2$ & 2                    \\
		\hline
		Kebnekaise K80  & 12GB $\times 2$ & 4                    \\
		\hline
		Kebnekaise V100 & 16GB            & 2                    \\
		\hline                
	\end{tabular}
	\label{table:proc-per-node}
\end{table}

%% file: results.tex
\section{Results}
\label{sec:results}
The goal of this work is to evaluate the potential and scalability of distributed TensorFlow for traditional HPC applications running on supercomputers with GPUs. We present the performance evaluation of the applications we designed and developed.

\subsection{STREAM Benchmark}

\begin{figure}[t]
	\begin{center}
		\includegraphics[width=0.7\linewidth]{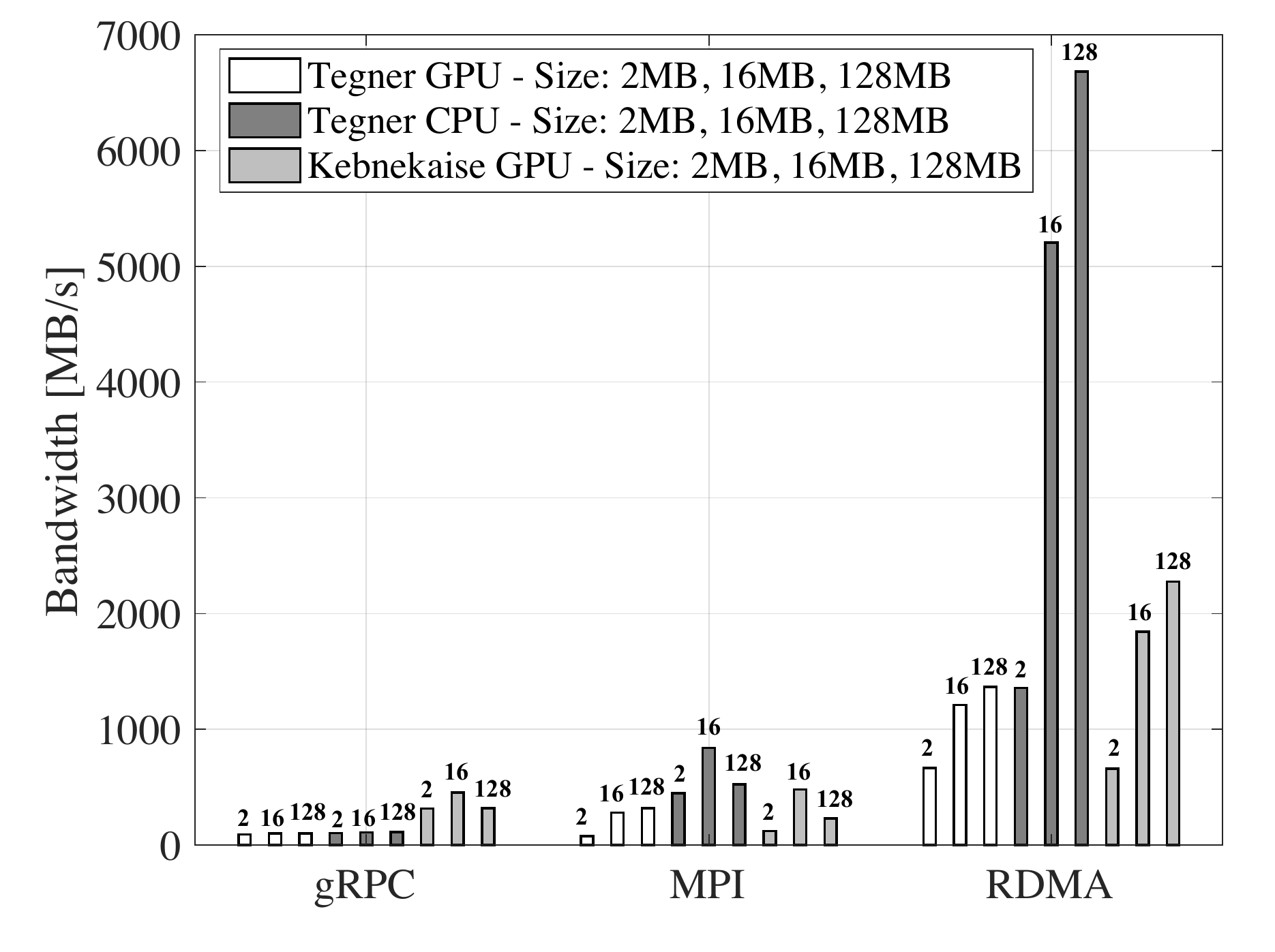}
		\caption{Communication performance in MB/s between two computing nodes on Tegner and Kebnekaise.}
		\label{fig:stream-bandwidth}
	\end{center}
\end{figure}

High performance interconnect is an important component of HPC system which supports synergy of parallel applications. Our testing shows that TensorFlow's RDMA module can take advantage of these infrastructures to transfer tensors residing on computing nodes. We evaluate communication performance with our STREAM micro-benchmark tool for three protocols: MPI, InfiniBand Verbs (RDMA) and gRPC. In each test, we invoke communication 100 times to create a stream of transfer and report average MB/s. We vary transfer size from two to 128 MB. We perform the tests with K420 nodes on Tegner and K80 nodes on Kebnekaise. On Tegner we test for when tensors reside in host and GPU memories. We specifically note that we use the default transfer configuration by TensorFlow MPI module for the tests with MPI. It means that instead of performing direct transfer of data to and from GPUs, tensors are first copied and serialized to host memory before transfer. This is due to that GPU Direct is not supported on our testing platforms.

Our results suggest that RDMA provides the best communication performance and we record peak bandwidth of over 6 GB/s on Tegner when tensors are placed in CPU host memory. The theoretical bandwidth on Tenger is 12 GB/s, which represents more than 50\% of bandwidth utilization. Fig.~\ref{fig:stream-bandwidth} shows the estimated communication performance in MB/s on Tegner and Kebnekaise. When using RDMA, bandwidth saturates at approximately 1300 MB/s on Tegner where tensors are hosted on K420 GPUs; on Kebnekaise, bandwidth saturates at below 2300 MB/s where tensors are hosted on K80 GPUs.

When communicating through MPI, bandwidth is less satisfactory. We measure approximately 318 MB/s on Tegner when tensors are hosted on K420 GPUs and MPI is used for communication; on Kebnekaise, we measure approximately 480 MB/s when tensors are hosted on K80 GPUs. One reason is due to copy and serialization process between GPU, host memory and inter-node transfer.

Finally, we show that gRPC gives the lowest bandwidth on Tegner. We note that this is due to gRPC connection is resolved to communicate through Ethernet. On Kebnekaise communicating through gRPC gives similar bandwidth to that of MPI.

\subsection{Tiled Matrix Multiplication} 

\begin{figure}[t]
	\begin{center}
		\includegraphics[width=0.7\linewidth]{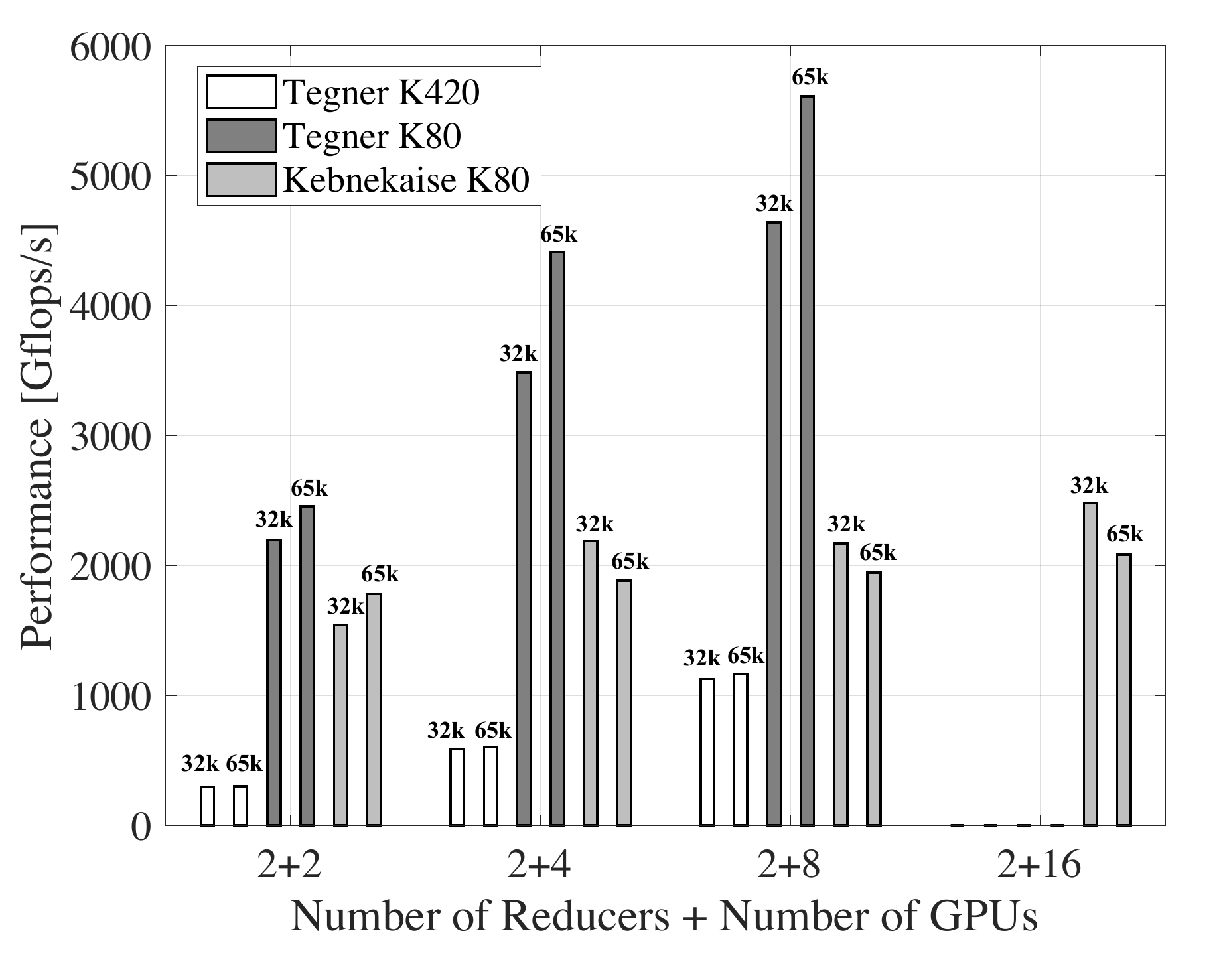}
		\caption{Performance of tiled matrix multiplication with different GPUs on Tegner and Kebnekaise in Gflops/s for different problem sizes.}
		\label{fig:matmul-gflops}
	\end{center}
\end{figure}

Our tiled matrix multiplication is implemented as a communication intensive application. Matrix tiles and results continuously flow through GPUs and the network. We show that the application gives good scalability on one of our test platform but gives sub-optimal scaling on another. We evaluate the performance of the application with three problem sizes: $16384 \times 16384$ ($2^{14}$), $32768 \times 32768$ ($2^{15}$) and $65536 \times 65536$ ($2^{16}$). The problems consist of randomly generated dense matrices. The matrices are pre-processed into tiles with tile size $4096 \times 4096$ and $8192 \times 8192$.

We test the application on Tegner and Kebnekaise. On Tegner we perform tests with two to eight K420 and K80 GPUs and on Kebnekaise we test with two to 16 K80 GPUs. To increase utilization, we use tile size $4096 \times 4096$ for K420 and run all problem sizes. For K80, we use tile size $8192 \times 8192$ and only run problem size $32768 \times 32768$ and $65536 \times 65536$. We run strong scaling tests by varying the number of GPUs being used and report results in Gflops/s. We estimate the flop count as $2N^{3}-N^{2}$. In our implementation we use two reducers to accumulate tiles with target index being odd number and even number.

We observe good scaling on Tegner. Fig.~\ref{fig:matmul-gflops} shows the performance results for tests conducted on different platforms. We measure approximately $2\times$ increase in performance when increasing the number of GPUs from two to four with K420 GPUs on Tegner for problem size $32768 \times 32768$. We observe similar performance improvement for this setting when increasing the number of GPUs in use from four to eight. For K80 GPUs on Tegner, we obtain roughly $1.8\times$ improvement when scaling from two to four GPUs with problem size $65536 \times 65536$.

We perform similar tests on K80 nodes on Kebnekaise. The scaling result is however less satisfactory. We measure peak performance of 2478 Gflops/s when running on 16 K80 GPUs for problem size $32768 \times 32768$. We obtain scaling of $1.4\times$ when scaling from two to four GPUs for the same problem size.

\begin{figure}[t]
	\begin{center}
		\includegraphics[width=0.5\linewidth]{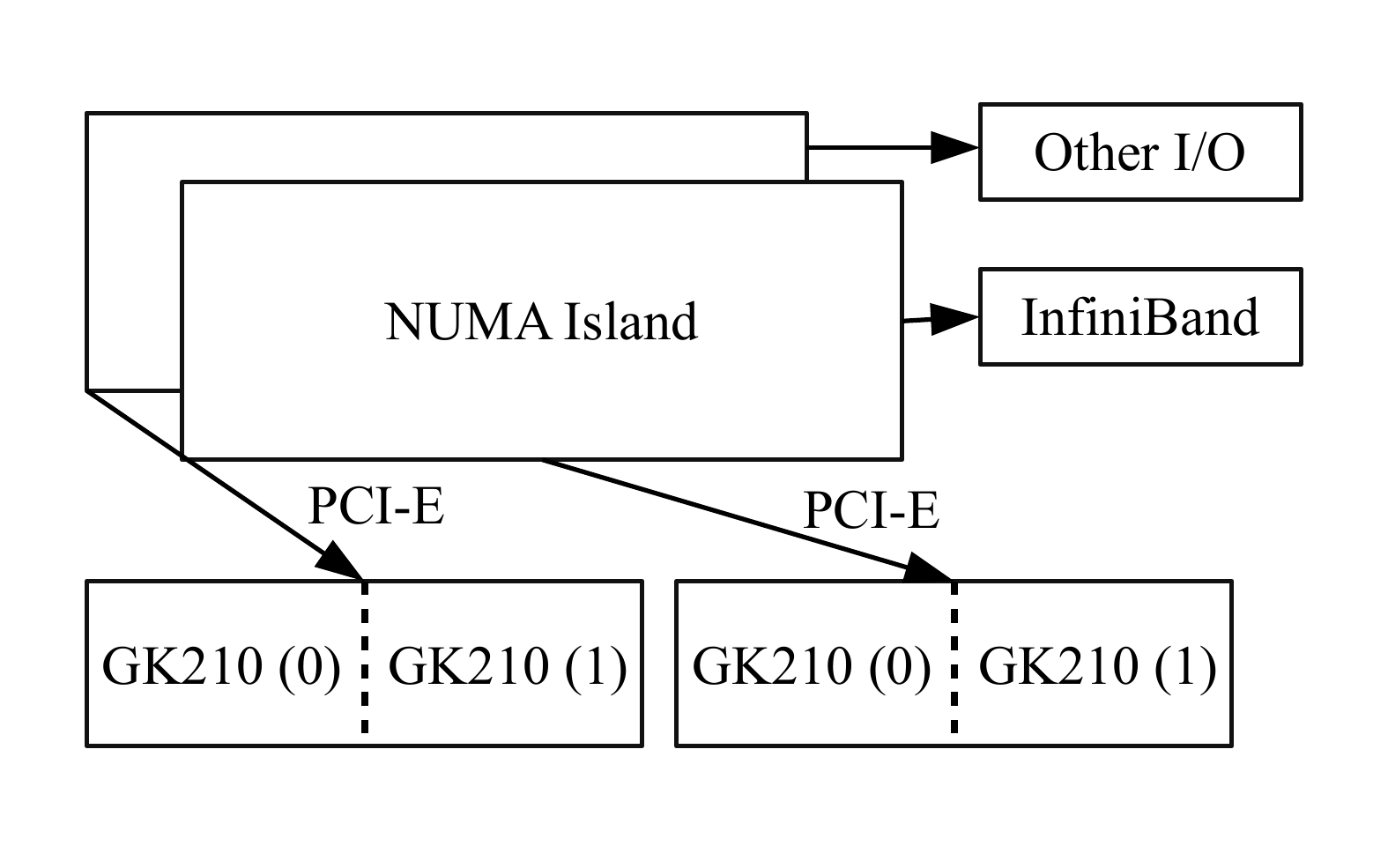}
		\caption{Topology of a GPU node on Kebnekaise}
		\label{fig:keb-topology}
	\end{center}
\end{figure}

To try understanding why scaling is sub-optimal on Kebnekaise we look at the configuration topology of its computing node. One possible explanation is due to the data-driven nature of the application where matrices are constantly flowing in and out of the systems. Fig.~\ref{fig:keb-topology} shows the topology of a GPU node of Kebnekaise\footnote{https://www.hpc2n.umu.se/resources/hardware/kebnekaise} where two K80 GPUs are located on two NUMA islands while I/O and network communication are only connected to either one island. With four instances of TensorFlow running per node comparing to only one or two instances on Tegner potentially means that the amount of data flowing and competing for bandwidth can be large, thus creating bottlenecks. 

\subsection{Conjugate Gradient Solver}

\begin{figure}[t]
	\begin{center}
		\includegraphics[width=0.7\linewidth]{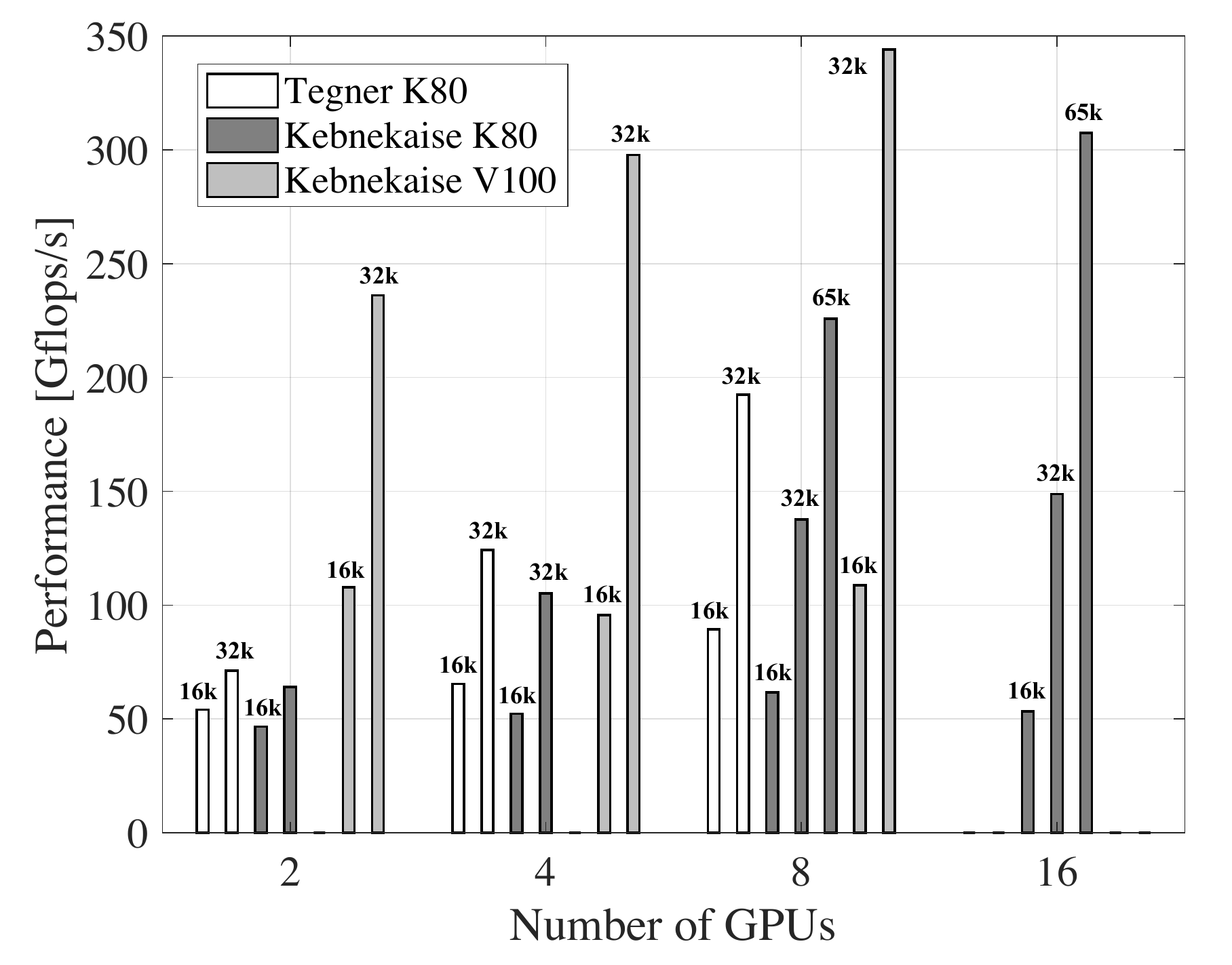}
		\caption{Performance of CG solver with problem size $16384 \times 16384$, $32768 \times 32768$ and $65536 \times 65536$ on Tegner and Kebnekaise with K420, K80 and V100 GPUs. Results are reported in Gflops/s.}
		\label{fig:cg-gflops}
	\end{center}
\end{figure}

Our CG solver exhibits generally good scaling across all of our testing platforms. We test the application on Tegner and Kebnekaise. On Tegner we test from two to eight K80 GPUs; on Kebnekaise we test with two to 16 K80 GPUs and with two to eight V100 GPUs. We run strong scaling tests for three problem sizes: $16384 \times 16384$ ($2^{14}$), $32768 \times 32768$ ($2^{15}$), $65536 \times 65536$ ($2^{16}$). The matrices are pre-processed by splitting into tiles where the size depends on the number GPUs used. For K80 GPUs on Tegner and V100 GPUs on Kebnekaise, we do not report result for problem size $65536 \times 65536$ due to insufficient memory and insufficient number of GPUs available; for K80 GPUs on Kebnekaise, we only report results for problem size $65536 \times 65536$ from eight to 16 GPUs. We report performance in Gflops/s and estimate flop count to be $500 \times 2 \times N^{2}$, where 500 is the number of iterations we run per test and $N^{2}$ belongs to run time dominating matrix vector multiplication.

The test results are summarized in Fig.~\ref{fig:cg-gflops}. We are only able to obtain little scaling for problem size $16384 \times 16384$ across different platforms. This is likely due to underutilization of GPUs after splitting the problem. This is particular obvious for tests on V100 GPUs on Kebnekaise as it is a powerful GPU. We observe a scaling of $1.6\times$ in performance when increasing from two to four K80 GPUs on Kebnekaise with problem size $32768 \times 32768$. When increasing the number of K80 GPUs in use from four to eight for the same setting, scaling drops to $1.3\times$, which is consistent with the expected behaviour of strong scaling. Similarly, we observe improvement of $1.36\times$ when scaling from eight to 16 K80 GPUs. Tests on V100 nodes on the other hand give $1.26\times$ improvement in performance when scaling the number of GPUs from two to four with problem size $32768 \times 32768$. When increasing the number of GPUs from four to eight improvement drops to $1.16\times$, indicating ratio between problem size and computation power must be optimized. We obtain similar scaling result on Tegner. We measure an approximately $1.74\times$ improvement in performance when scaling from two to four K80 GPUs with problem size $32768 \times 32768$.

Comparing to the performance of the matrix-matrix multiplication application, the CG solver gives better scaling on both Tegner and Kebnekaise. A key difference is that the amount of data flowing between GPUs and reducer is relatively little, consisting mostly of vectors and scalars instead of dense matrices.

Despite the difficulty in direct comparison against other frameworks, we refer to several related works on CG solver implementation on GPU for give an overview of related performance results. A similar implementation of task-based CG algorithm was implemented in Ref.~\cite{10.1007/978-3-319-58943-5_6} with StarPU on a system with three NVIDIA Tesla M2070 GPUs and they recorded performance close to 30 Gflops/s on three GPUs. Ref.~\cite{markidis2015openacc} ports and evaluates NekBone, a computational fluid dynamics mini-application to GPU with OpenACC which relies on a CG solver. They evaluated the implementation on a cluster with NVIDIA K20 GPUs and reached up to 43 Gflops/s on a single node. Our CG solver, running on eight V100 GPUs gave over 300 Gflops/s.

\subsection{Fast Fourier Transform}
The scaling performance of our FFT application is overall satisfactory. We test the application on Tegner with both K420 and K80 GPUs. For K420, we evaluate problem size $2^{29}$ with 64 tiles with size $2^{23}$; for K80, we evaluate problem size $2^{31}$ with 128 tiles with size $2^{24}$. We perform strong scaling tests and scale from two to eight GPUs. Since we perform serial merging of results in Python with one process, the cost is theoretically constant regardless of the number of GPUs used. For this reason, we only report scaling results from the beginning of the application to when all tiles are collected by the merger. Results are reported in Gflops/s and we estimate the flop count of the application with $5 \times N \log_{2}{N}$.

\begin{figure}[t]
	\begin{center}
		\includegraphics[width=0.7\linewidth]{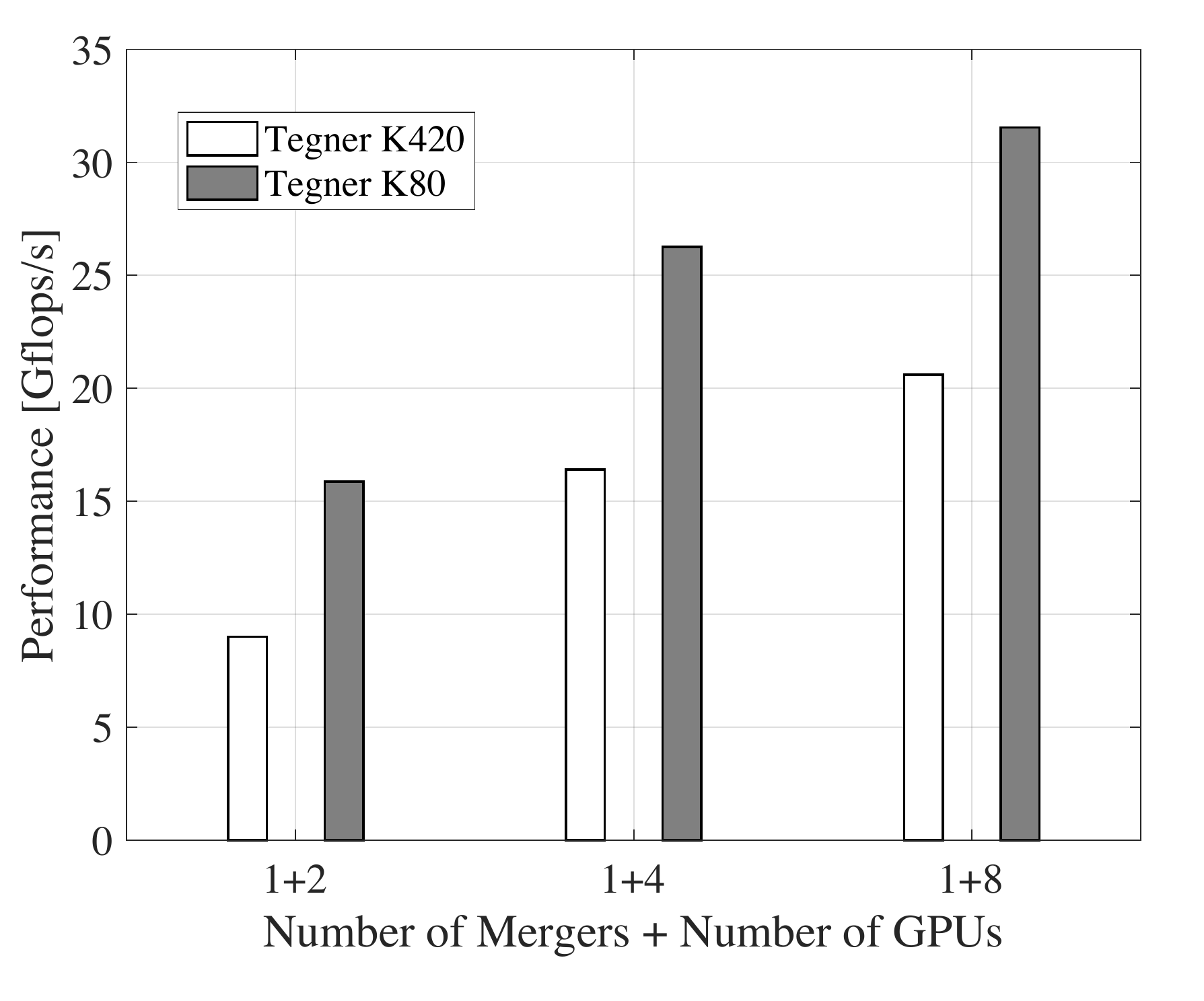}
		\caption{Performance of FFT solver on Tegner with problem size $2^{31}$ in 128 tiles of problem size $2^{29}$ with K80 GPUs in Gflops/s; problem size $2^{29}$ in 64 tiles of size $2^{23}$ with K420 GPUs in Gflops/s.}
		\label{fig:fft-flops}
	\end{center}
\end{figure}

We observe good scaling when increasing the number of GPUs in use from two to four for both GPUs. The results are reported in Fig.~\ref{fig:fft-flops}. When increasing from two to four GPUs for both configurations there is an approximately $1.6\times$ to $1.8\times$ increase in performance. However when increasing from four to eight GPUs the performance improvement clearly flattens out. One reason is underutilization of GPUs when the number of tiles processed per GPU decreases.

%% file: related-work.tex
\section{Related Work}
\label{sec:related-work}

The TensorFlow programming framework was released open-source in 2015 and presented in a seminal paper by \emph{Abadi et al.}~\cite{tensorflow} at the USENIX Symposium on Operating Systems Design and Implementation in 2016. This paper presents the basic TensorFlow concepts and design. A hands-on description of how to use TensorFlow in distributed environment is included in the book~\cite{geron2017hands}. The performance evaluation of TensorFlow under ML workloads was presented in Ref.~\cite{chien2018characterizing} and Ref.~\cite{mathuriya2018cosmoflow} studied TensorFlow application running on supercomputers.

In terms of communication performance, Ref.~\cite{grpc-benchmark} designed a benchmark suite to study the communication performance of TensorFlow. Refs.~\cite{tensorflow-mpi} and \cite{horovod} investigated the use of MPI collectives for communication between TensorFlow processes. Studies have also been done to investigate different implementation of RDMA on TensorFlow to improve communication efficiency~\cite{tensorflow-rdma-performance}\cite{xue2018rpc}\cite{tensorflow-rdma-zero-copy}.

The main competitor of TensorFlow is PyTorch~\cite{paszke2017automatic}, an open-source Python framework. PyTorch supports execution of application on distributed systems with GPUs. The main difference with TensorFlow is that PyTorch is based on eager execution model instead of the current TensorFlow default deferred execution model. As it is easier to program and debug code in eager mode, an increasing number of users begins to use PyTorch. However, it is likely that future release of TensorFlow will have eager execution mode by default.

Some other deep-learning frameworks that are increasingly being adopted in HPC systems include Caffe~\cite{jia2014caffe}. Caffe focuses on GPU training on a single node. S-Caffe extends Caffe and provides distributed GPU training on a cluster through co-designing the framework with CUDA-aware MPI to provide better interoperability between CUDA and MPI~\cite{s-caffe}.

HPC frameworks that are using computational graph and are similar to TensorFlow despite not designed for ML workloads, are PaRSEC (Parallel Runtime Scheduling and Execution Controller) and StarPU. PaRSEC~\cite{parsec} is a programming framework for distributed many-core heterogeneous architectures. Applications in PaRSEC are expressed as a Direct Acyclic Graph (DAG) of tasks with edges representing dependencies. StarPU~\cite{starpu} is a task-based programming framework where tasks are represented as a series of DAG.

%% file: conclusion.tex
\section{Discussion and Conclusion}
\label{sec:conclusion}

In this paper we introduced the development paradigm of HPC application with TensorFlow. We designed and implemented four common HPC applications and performed experiments on different HPC platforms with GPUs. We also introduced and discussed the distributed programming model of TensorFlow. TensorFlow currently supports a parameter server-worker model where workers perform individual work with new data and update parameter. This however, presents a challenge when developing HPC applications that are based on domain decomposition. In addition, this hampers the scalability of large scale deployment with large number of machines. Some of the efforts to increase support to these applications include Uber's Horovod~\cite{horovod} and Cray's Machine Learning Plugin\footnote{https://www.cray.com/products/analytics/urika-xc}~\cite{mathuriya2018cosmoflow}. These plugins enable the development of application with MPI like interfaces through an MPI communication backend for functions such as \emph{allreduce} without needing the use of dedicated servers for parameters.

During our development, one difficulty is the reformulation of algorithms that require reduction, where we have to implement our reducer with queue dataflow mechanisms. Another limitation is imposed by Python. One example is the Global Interpreter Lock which prevents concurrent thread execution, which QueueRunners are dependent on. Another example is the Python's relatively low performance in numerical computation. This hampers performance of applications where logic are difficult to express in computation graph or applications that are partly programmed with TensorFlow and partly in Python. An example is our FFT application. The process of merging in Python takes considerably longer execution time then the computation part by TensorFlow. In fact we initially discovered that directly performing slicing insertion into a local \emph{Numpy} array during the extraction of tiles from the graph already hampers overall performance thus preventing any scaling during the computation of FFT on tiles by TensorFlow.

Despite all of that, TensorFlow is a rapidly growing programming framework that can be used for  development of HPC applications on supercomputers with accelerators. It provides a complete programming environment ranging from high-level APIs to a C++ runtime for device management and communication. The high level APIs allow easy development and deployment of distributed algorithms without needing in-depth knowledge into concepts such as CUDA and MPI. In fact, our distributed CG solver with checkpoint-restart capability only consists of less than 300 lines of code and our matrix multiplication application consists only of less than 240 lines of code. TensorFlow's deferred execution model presents an opportunity for runtime optimization and improved auto parallelization.

Our performance results demonstrated that TensorFlow is a promising framework. All of our applications which are formulated with data-driven approach show good scaling when the number of GPUs being used increases, without requiring the use of low level APIs such as CUDA and MPI. We showed that TensorFlow has high potential to become a HPC programming framework for heterogeneous supercomputers.
